\documentclass[12pt,preprint]{aastex}

\begin{document}

\def\cfa{1}
\def\nrao{2}
\def\ias{3}
\def\ef{4}
\def\jansky{5}
\def\ta{6}
\def\taka{7}
\def\takaa{8}
\def\tokyo{9}
\def\ssl{10}
\def\nasa{11}
\def\uva{12}
\def\ucb{13}
\def\su{14}
\def\bonn{15}
\def\york{16}
\def\hart{17}
\def\russia{18}
\def\alb{19}
\def\usra{20}
\def\mpi{21}
\def\ru{22}
\def\emi{23}

\title{Panchromatic Observations of SN\,2011dh Point to a Compact Progenitor Star}

\author{A.~M. Soderberg\altaffilmark{\cfa},
  R. Margutti\altaffilmark{\cfa}, B.~A. Zauderer\altaffilmark{\cfa},
  M. Krauss\altaffilmark{\nrao}, B. Katz\altaffilmark{\ias,\ef},
  L. Chomiuk\altaffilmark{\cfa,\nrao,\jansky},
  J.~A. Dittmann\altaffilmark{\cfa}, E. Nakar\altaffilmark{\ta},
  T. Sakamoto\altaffilmark{\taka,\takaa},
  N. Kawai\altaffilmark{\tokyo}, K. Hurley\altaffilmark{\ssl},
  S. Barthelmy\altaffilmark{\nasa}, T. Toizumi\altaffilmark{\tokyo},
  M. Morii\altaffilmark{\tokyo}, R.~A. Chevalier\altaffilmark{\uva},
  M. Gurwell\altaffilmark{\cfa}, G. Petitpas\altaffilmark{\cfa},
  M. Rupen\altaffilmark{\nrao}, K.~D. Alexander\altaffilmark{\cfa},
  E.~M. Levesque\altaffilmark{\ucb}, C. Fransson\altaffilmark{\su},
  A.~Brunthaler\altaffilmark{\bonn},
  M.~F. Bietenholz\altaffilmark{\york,\hart},
  N. Chugai\altaffilmark{\russia}, J.~Grindlay\altaffilmark{\cfa}, A. Copete\altaffilmark{\cfa},
  V. Connaughton\altaffilmark{\alb}, M. Briggs\altaffilmark{\alb}, C. Meegan\altaffilmark{\usra}, A. von Kienlin\altaffilmark{\mpi}, X. Zhang\altaffilmark{\mpi}, A. Rau\altaffilmark{\mpi} S. Golenetskii\altaffilmark{\ru}, E. Mazets\altaffilmark{\ru}, T. Cline\altaffilmark{\emi}}

\altaffiltext{\cfa}{Harvard-Smithsonian Center for Astrophysics, 60 Garden St., Cambridge, MA 02138, USA}
\altaffiltext{\nrao}{National Radio Astronomy Observatory, Socorro, NM 87801, USA}
\altaffiltext{\ias}{School of Natural Sciences, Institute for Advanced Study, Einstein Drive, Princeton, NJ 08540, USA}
\altaffiltext{\ef}{Bachall Fellow, Einstein Fellow}
\altaffiltext{\jansky}{Chomiuk is a Jansky Fellow of the National
  Radio Astronomy Observatory}
\altaffiltext{\ta}{Raymond and Beverly Sackler School of Physics \& Astronomy, Tel Aviv University, Tel Aviv 69978, Israel}
\altaffiltext{\taka}{Center for Research and Exploration in Space Science and Technology (CRESST), NASA Goddard Space Flight Center, Greenbelt, MD 20771, USA}
\altaffiltext{\takaa}{Department of Physics, University of Maryland, Baltimore County, 1000 Hilltop Circle, Baltimore, MD 21250, USA}
\altaffiltext{\tokyo}{Department of Physics, Tokyo Institute of Technology, 2-12-1 Ohokayama, Meguro, Tokyo 152-8551, Japan}
\altaffiltext{\ssl}{Space Sciences Laboratory, University of California, 7 Gauss Way, Berkeley, CA 94720-7450, USA}
\altaffiltext{\nasa}{NASA Goddard Space Flight Center, Greenbelt, MD 20771, USA}
\altaffiltext{\uva}{University of Virginia, Astronomy Department, Charlottesville, VA 22904, USA}
\altaffiltext{\ucb}{CASA, Dept. of Astrophysical and Planetary Sciences, University of Colorado 389-UCB, Boulder, CO 80309, USA}
\altaffiltext{\su}{Department of Astronomy, The Oskar Klein Centre, Stockholm University, 106 91 Stockholm, Sweden}
\altaffiltext{\bonn}{Max-Planck-Institute f\"ur extraterrestrische Physik,
Giessenbachstra{\ss}e,  85748 Garching, Germany}
\altaffiltext{\york}{Dept. of Physics and Astronomy, York University, Toronto, M3J 1P3, Ontario, Canada}
\altaffiltext{\hart}{Hartebeesthoek Radio Observatory, P.O. Box 443, Krugersdorp, 1740, South Africa}
\altaffiltext{\russia}{Institute of Astronomy, RAS, Pyatnitskaya 48, Moscow 11917, Russia}
\altaffiltext{\alb}{Physics Department, The University of Alabama in Huntsville, Huntsville, AL 35809, USA}
\altaffiltext{\usra}{Universities Space Research Association, NSSTC, 320 Sparkman Drive, Huntsville, AL 35805, USA}
\altaffiltext{\mpi}{Max-Planck-Institut für extraterrestrische Physik, Giessenbachstrasse 1, 85748 Garching, Germany}
\altaffiltext{\ru}{Ioffe Physical-Technical Institute of the Russian Academy of Sciences, St. Petersburg 194021, Russia}
\altaffiltext{\emi}{Emeritus, NASA Goddard Space Flight Center, Code 661, Greenbelt, MD 20771, USA}

\begin{abstract}
  We report the discovery and detailed monitoring of X-ray emission
  associated with the Type IIb SN\,2011dh using data from the {\it
    Swift} and {\it Chandra} satellites, placing it among the best
  studied X-ray supernovae to date. We further present millimeter and
  radio data obtained with the SMA, CARMA, and EVLA during the first
  three weeks after explosion.  Combining these observations with
  early optical photometry, we show that the panchromatic dataset is
  well-described by non-thermal synchrotron emission (radio/mm) with
  inverse Compton scattering (X-ray) of a thermal population of
  optical photons.  In this scenario, the shock partition fractions
  deviate from equipartition by a factor,
  $(\epsilon_e/\epsilon_B)\sim 30$.  We derive the properties of the
  shockwave and the circumstellar environment and find a time-averaged
  shock velocity of $\overline{v}\approx 0.1c$ and a progenitor mass
  loss rate of $\dot{M}\approx 6\times 10^{-5}~\rm M_{\odot}~yr^{-1}$
  (for an assumed wind velocity, $v_w=1000~\rm km~s^{-1}$).  We show
  that these properties are consistent with the sub-class of Type IIb
  supernovae characterized by compact progenitors (Type cIIb) and
  dissimilar from those with extended progenitors (Type eIIb).
  Furthermore, we consider the early optical emission in the context
  of a cooling envelope model to estimate a progenitor radius of
  $R_*\approx 10^{11}$ cm, in line with the expectations for a Type
  cIIb supernova.  Together, these diagnostics are difficult to
  reconcile with the extended radius of the putative yellow supergiant
  progenitor star identified in archival {\it HST} observations,
  unless the stellar density profile is unusual.  Finally, we searched
  for the high energy shock breakout pulse using X-ray and gamma-ray
  observations obtained during the purported explosion date range.
  Based on the compact radius of the progenitor, we estimate that the
  shock breakout pulse was detectable with current instruments but
  likely missed due to their limited temporal/spatial coverage. Future
  all-sky missions will regularly detect shock breakout emission from
  compact SN progenitors enabling prompt follow-up observations with
  sensitive multi-wavelength facilities.

\end{abstract}


\section{Introduction}
\label{sec:intro}

A key open question in the study of core-collapse supernovae (SNe) is
the nature and diversity of their progenitor systems.  High-resolution
optical imaging of nearby galaxies has firmly established that the
progenitors of Type IIP SNe are red supergiants
\citep{sma09}, while pre-discovery imaging for Type II SNe 1987A
\citep{gcc+87}, 1993J \citep{ahr94} and 2005gl \citep{glf+07,gl09}
point to a diverse set of massive stars. An alternative route to the
nature of the progenitors is to obtain panchromatic follow-up
observations within days of explosion.  The shock breakout pulse and
subsequent adiabatic cooling of the ejecta can yield information on
the progenitor size since the duration and energy of these signals
scales with the size of the progenitor star
\citep{col74,eb92,wmc07,cf08,kbw10,ns10}.  Complementary follow-up
observations at radio, millimeter, and X-ray bands provide unique
diagnostics on the shockwave velocity which scale inversely with the
progenitor radius.  The utility of such a multi-wavelength technique
was demonstrated by the serendipitous X-ray discovery and
comprehensive follow-up study of SN\,2008D \citep{sbp+08,cf08}.

On May 31.893 UT amateur astronomer Amadee Riou discovered an optical
transient in M51 ($d=8.4\pm 0.6$ Mpc;
\citealt{fcj97} see also \citealt{vinko11}).
%  Its a Seyfert 2;     NGC 5194        UGC 8493  4C 47.36A 
%  FeldmeierCJ1997 says Dm = 29.62 +- 0.1 => 8.4 +- 0.4 Mpc
Multiple individuals and groups subsequently confirmed the transient
using pre- and post-discovery imaging \citep{gmg+11,sfc11}.  The
Palomar Transient Factory (PTF; \citealt{lkd+09}) reported a deep
non-detection in pre-discovery data constraining the onset of the
optical emission to be May 31.275-31.893 UT
\citep{agy+11}.  Based on an initial spectrum on June 3.3 UT, the transient was
classified as a Type II supernova, designated SN\,2011dh 
\citep{sfc11}. Further spectroscopy revealed evidence for helium
absorption features prompting the re-classification as Type IIb
\citep{agp+11,mkw+11}.

A putative progenitor star has been identified in pre-explosion {\it
Hubble Space Telescope} (HST) images with a spectral energy
distribution consistent with a yellow supergiant
\citep{mfe+11,vlc+11}.  The mass of the star is estimated to be
between $M_{ZAMS}\approx 13$ and $21~M_{\odot}$, but
temperature-dependent bolometric corrections, discrepancies between
evolutionary tracks, and treatments of rotation should also be
carefully considered \citep{dmm+09}.  Based on the estimated
luminosity and temperature of the object, the stellar radius is
$R_*\approx 10^{13}$ cm \citep{ph11}.  Both \citet{mfe+11} and \citet{vlc+11} and
discuss the possibility that the yellow supergiant is
instead the binary companion to the SN\,2011dh progenitor star.  In this
scenario, the actual progenitor star would have had a smaller radius.

Recently, \citet{cs10} proposed that SNe IIb may be divided into two
sub-classes based on the radius and mass loss history of the
progenitor star and the properties of the shockwave.  In this
framework, compact progenitors ($R_*\sim 10^{11}$ cm) with modulated
radio light-curves and shockwave velocities of
$\overline{v}\sim 0.1$c are identified as SNe cIIb with members
including SNe 2001ig \citep{rss+04}, 2003bg \citep{sck+06}, and 2008ax
\citep{rpb+09}.  Meanwhile, extended progenitors ($R_*\sim 10^{13}$
cm) with smooth radio light-curves and slower shockwaves are
identified as SNe eIIb (e.g., SN\,1993J; \citealt{bbr+02,wwp+07}).  The
modulated radio emission points to an unusual
(perhaps episodic) mass loss that may be unique to SNe cIIb.

SN\,2011dh showed an initial peak magnitude of $M_g\approx -16.5$ mag
at $\Delta t\approx 1$ day since explosion before fading quickly
\citep{agy+11,psp+11}.  The SN then re-brightened at $\Delta t\approx
5$ days.  The two light-curve components may be interpreted as cooling
envelope emission followed by a rebrightening due to the
radioactive decay of $^{56}$Ni.  Based on a comparison of the
SN\,2011dh light-curve with that of SN\,1993J and a photospheric
temperature measurement, \citet{agy+11} proposed that SN\,2011dh
belongs to the Type cIIb class.

Here we report the discovery and monitoring of X-ray emission
associated with SN\,2011dh and present radio and mm-band detections
from the first few weeks after explosion. We show that the radio and
X-ray properties are consistent with those of SNe cIIb and dissimilar
from those of SNe eIIb suggesting that the progenitor was compact at
the time of explosion.  This is supported by our modeling of the early
cooling envelope emission, which points to a progenitor radius of
$R_*\approx 10^{11}$ cm.  Together, these diagnostics suggest that the
putative yellow supergiant progenitor is instead a binary companion or
unrelated to the SN.  Finally we present a detailed compilation of
X-ray and gamma-ray observations from multiple satellites and
instruments obtained during the purported explosion date range.  We
estimate that the shock breakout pulse was detectable with current
high energy instruments but likely missed due to their limited
temporal/spatial coverage.  

\section{Observations}
\label{sec:obs}

Following the optical discovery of SN\,2011dh, we initiated a prompt
panchromatic follow-up campaign to map the non-thermal properties of
the ejecta.

\subsection{Swift/XRT Observations} 
\label{sec:xrt}

\begin{figure}
\plotone{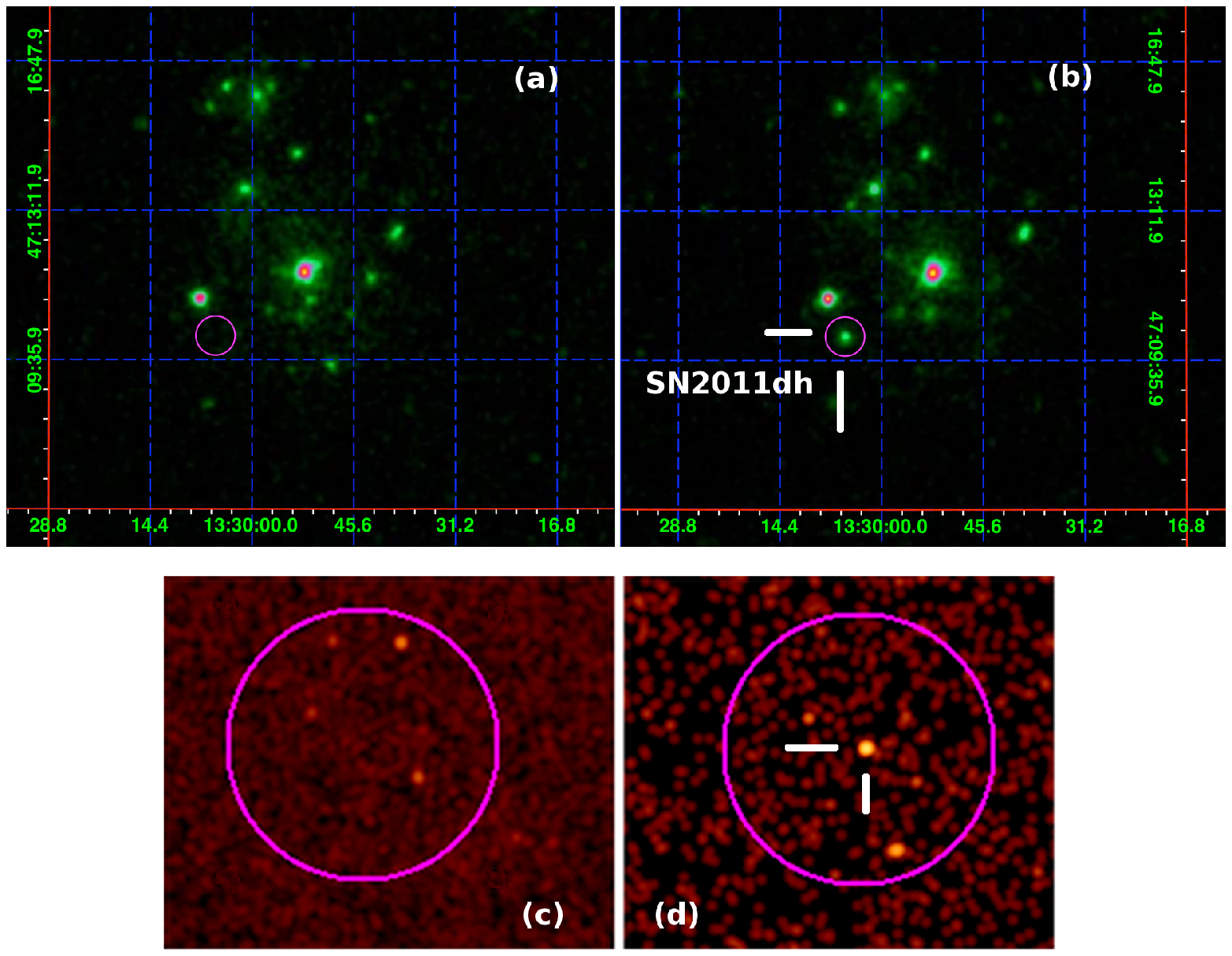}
\caption{{(a)}: A (0.3-10) keV co-added archival image 
(total exposure of 64 ks) of M51 obtained pre-explosion using {\it
Swift}/XRT in the time range June 2006 - May 2009.  No statistically
significant excess is detected at the SN2011dh position. {(b)}: XRT
observations of M51 (total exposure of 78 ks) obtained after the
discovery of SN\,2011dh clearly reveal an X-ray source at the SN
position.  {(c)}: Archival {\it Chandra} observations reveal no source
within 28 arcsec of the SN position (circle). {(d)}: SN\,2011dh is
detected in our {\it Chandra} ToO observation on July 3.
No bright contaminating X-ray sources are detected within the
extraction region. We restrict our XRT extraction region at
late times to avoid the faint source to the SW.}
\label{fig:image}
\end{figure}

Thanks to multiple Target-of-Opportunity requests, {\it Swift}/XRT
promptly observed SN\,2011dh on June 3.50, just $\sim 3$ days after the
explosion. As initially reported in \citet{ms11}, we discovered a
bright X-ray source (S/N$\sim10$) at coordinates,
$\rm{RA_{\rm{J2000}}}=13^{\rm{h}}30^{\rm{m}}5.18^{\rm{s}}$,
$\rm{Dec_{J2000}}=+47^{\circ}10'11.14''$ (uncertainty $4''$ radius,
90\% confidence) at $1''$ from the optical SN position.  This data analysis
reported here supersedes the preliminary analysis presented in the circulars; we 
find a count-rate of $\sim0.015\,\rm{cps}$ for this first epoch. We analyzed 69 ks of archival
pre-SN {\it Swift}/XRT observations and these data reveal no X-ray
source at the SN position with a $3\,\sigma$
upper limit of $5.6\times10^{-4}\rm{cps}$. This fact, coupled to the
spatial coincidence of the SN, strongly suggests that the new source
represents the X-ray counterpart to SN\,2011dh
(Figure~\ref{fig:image}).

Observations of SN\,2011dh with {\it Swift} continued for the next
several weeks.  We retrieved and analyzed the XRT data from the
HEASARC archive collected in the time period, June 3 - July 3 UT
(total exposure time of $137\,\rm{ks}$).  All XRT data were analyzed
with the \textsc{heasoft} (version 6.10) software package and
corresponding calibration files; standard filtering and screening
criteria were applied.  Due to the proximity of a nearby, steady X-ray
source (Figure~\ref{fig:image}), we adopted a 12-pixel ($\sim 28''$)
extraction region centered on the optical position; for lower count
rates ($< 0.0025$ cps) we reduced the extraction region to a radius of
6-pixels to increase the S/N ratio and eliminate contamination from a
nearby faint source.  The background was estimated from the
pre-explosion {\it Swift}/XRT data to properly account for the
contamination from the extended X-ray emission associated with M51.  A
spectrum extracted over June 3-17 UT can be modeled by an absorbed
power-law with photon index $\Gamma=1.5\pm0.2$ (90\% c.l.)  assuming a
Galactic foreground column density
$\rm{N_H}=1.81\times10^{20}\,\rm{cm^{-2}}$ (\citealt{kbh+05}) and no
intrinsic absorption ($\chi^2/\rm{dof}=70.7/73$,
P-val=0.56). Alternatively, a thermal plasma spectral model with
best-fitting temperature $kT=7.5^{+9.7}_{-3.3}$ keV (90\% c.l.) can
adequately represent the data ($\chi^2/\rm{dof}=68.8/73$, P-val=0.62).
Both models give an average unabsorbed flux of $F_X\approx
1.65\times10^{-13}\,\rm{erg\,cm^{-2}\,s^{-1}}$ (0.3-8 keV)
corresponding to a luminosity, $L_X\approx 2\times
10^{39}\,\rm{erg\,s^{-1}}$.

Our resulting X-ray light-curve is shown in
Figure~\ref{Fig:2011dhvs1993J} and Table~\ref{tab:xray}.  Over the first 10 days the SN faded by
a factor $\approx8$. Adopting a simple power-law model for the decay,
we derive an index of $\alpha=-0.8\pm 0.2$ (90\% c.l.).  The source
shows some evidence for spectral softening with time, with the photon
index evolving from $\Gamma_1=0.9\pm0.3$ (90\% c.l., $\Delta t=3-7$
days) to $\Gamma_2=1.8\pm0.2$ (90\% c.l., $\Delta t=7-17$ days).
%This scenario is confirmed by the different rates of decay of the
%0.3-2.4 keV vs . 2-8 keV light-curves (Fig. \ref{Fig:lcbands}): 
In comparison with X-ray observations of SN\,1993J, the softening
observed for SN\,2011dh begins at an earlier epoch.  

\subsection{Chandra Observations}
\label{sec:chandra}

We supplement our {\it Swift}/XRT light-curve with two {\it Chandra}
observations.  As reported by \citealt{poo+11}, SN\,2011dh was
observed with {\it Chandra} for 10 ksec beginning on June 12.3 UT.  The
SN was detected with a count rate of $\sim 0.0115$ cps.
Adopting a power-law spectral model, they derive an X-ray flux of
$F_X=(1.0\pm 0.3)\times 10^{-13}~\rm erg~cm^{-2}~s^{-1}$ (0.5-8 keV).  

On July 3.4 UT, we obtained a second 10 ksec observation of SN\,2011dh
with the {\it Chandra} Advanced CCD Imaging Spectrometer (ACIS) under
a Target-of-Opportunity program (PI Soderberg).  Data were reduced
with the CIAO software package (version 4.3), with calibration
database CALDB (version 4.4.2).  We applied standard filtering using
CIAO threads for ACIS data.  We clearly detect a source at the SN
position.  Extracting within a 15 pixel aperture, we derive a
(background-subtracted) count rate of $\sim 0.0049$ cps.  Adopting the
same power-law spectral model described in \S\ref{sec:xrt}, we derive
an unabsorbed X-ray flux of $F_X=(2.75\pm 0.55)\times 10^{-14}~\rm
erg~s^{-1}~cm^{-2}$ (0.3-8 keV).  We further note that we do not detect any
bright nearby sources that would otherwise contaminate the extraction
region adopted for the {\it Swift}/XRT data.

In Table~1 and Figure~\ref{Fig:2011dhvs1993J} we report the combined {\it
Swift}/XRT and {\it Chandra} light-curve of SN\,2011dh, representing
the best-sampled X-ray light-curve for a SN IIb to date.  We compare
the X-ray properties of SN\,2011dh with those of other SNe IIb
including SNe 1993J \citep{cdr+09}, 2001gd \citep{pam+05}, 2008ax
\citep{rpb+09}, 2001ig \citep{sr02}, and 2003bg \citep{sck+06}.  As is
clear from the Figure, the SN\,2011dh X-ray light-curve is more
closely related to Type cIIb explosions and a factor of $\sim 10$ less
luminous than that observed for the Type eIIb SNe 1993J and 2001gd.
We  suggest that the Type cIIb sub-class {\it may} be further
characterized by low X-ray luminosities and early spectral
softening.

\subsection{CARMA Observations}
\label{sec:carma}

We observed SN\,2011dh with the Combined Array for Research in
Millimeter-wave Astronomy (CARMA; \citealt{bbh+06}) beginning on 2011
June 4.1 UT.  Observations were conducted with CARMA's nine 6.1-m
antennas and six 10.4-m antennas in the D configuration, with a
maximum baseline length of 150 m.  We implemented radio and optical
pointing \citep{cwc10}.  We selected central frequencies of
$\nu=107 $ GHz and 230 GHz, with a total bandwidth of $\sim$ 6 GHz.
%One sideband was tuned to
%observe CO(1-0) and (2-1), respectively, with higher velocity
%resolution.  
Gain calibration was performed with J1153+495 and we used a
source-calibrator cycle time of $\sim$ 15-20 minutes. Flux and
bandpass calibration was carried out using observations of 3C273 and 3C345
and Neptune resulting in an overall uncertainty in the absolute flux
calibration of $\sim 10\%$.  We used the Multichannel Image
Reconstruction Image Analysis and Display (MIRIAD; \citealt{stw95})
software package for data reduction.  We integrated on SN\,2011dh for
43 minutes at each frequency; a source is clearly detected at
$\nu=107$ GHz that is coincident with the optical and X-ray SN positions.  Preliminary results for the
$\nu=107$ GHz observation were presented by \citet{hzc11}. Here we
present the results of the $\nu=230$ GHz observation and
a re-analysis of the $\nu=107$ GHz data in which we have refined the flux
calibration. Fitting a Gaussian model to the source, we derive an
integrated flux density of $F_{\nu}=4.5\pm 0.3$ mJy at $\nu=107$ GHz
and a 3$\sigma$ upper limit of $F_{\nu}\lesssim 3.5$ mJy at $\nu=230$
GHz (Table~2).  Thus the radio spectrum is optically-thin between the two
mm-bands at $\Delta t\approx 4$ days since explosion.
 
\subsection{SMA Observations}
\label{sec:sma}

Contemporaneously with the CARMA observations, we observed SN\,2011dh
with the Submillimeter Array (SMA;~\citealt{hml04}) on June 4.0 UT in
the compact configuration at a frequency of $\nu=230$ GHz with 8~GHz
bandwidth.  Observations included all eight antennas.  Passband
calibration was performed in the standard way using Neptune, 3C454.3
and 3C279.  We used 3C279 for flux calibration, verifying our
calibration with observations of Titan in addition to 3C279 and our
gain calibrators on June 10 UT.  The absolute flux calibration is
accurate to $\sim$10\%.  We flagged low elevation data ($<$
21$^{\circ}$) and the first several hours of the observation when
weather conditions were less favorable, prior to improved  pointing
solutions, and when one antenna was missing.  The data were calibrated
using standard MIR/IDL routines developed for the SMA, with further
calibration and imaging carried out in MIRIAD and the Astronomical
Image Processing System (AIPS; \citealt{gre03}).  The resulting total
integration time on source was 2.75 hours. We detect a radio source
coincident with the SN position with a flux density of $F_{\nu}=3.6\pm
0.9$ mJy (Table~2).  This detection is not inconsistent with the $3\sigma$
upper limit from CARMA.

\subsection{EVLA Observations}
\label{sec:evla}

\begin{figure}
\epsscale{1.0}
\plotone{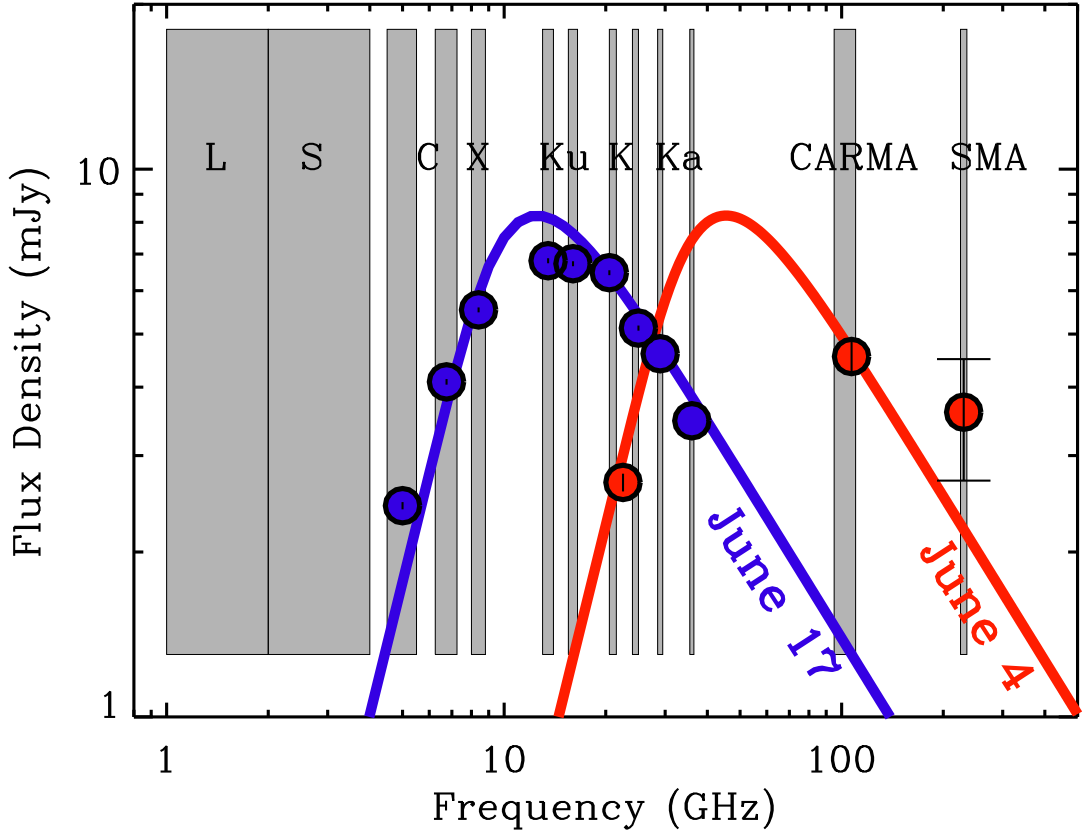}
\caption{The radio spectrum of SN\,2011dh across multiple epochs --  
$\Delta t\approx 4$ (red) and 17 (blue) days -- is well
described by a synchrotron self-absorbed spectral model with $F\propto \nu^{5/2}$ ($F_{\nu}\propto \nu^{-(p-1)/2}$) below (above) the spectral peak, $\nu_p$.
The observations indicate an electron energy index of $p\approx 3$.  Error
bars are 1$\sigma$.  The gray bands mark the EVLA, CARMA, and SMA
bands used in our long-term study of SN\,2011dh as the spectral peak cascades
to lower frequencies with time (see \citealt{krauss11} for a
detailed discussion).}
\label{fig:spectrum2}
\end{figure}

On June 4.25 UT, a radio counterpart was detected with the Expanded
Very Large Array (EVLA; \citealt{pnj+09}) with a flux density of
$F_{\nu}=2.68\pm 0.10~\rm mJy$ at $\nu=22.5$ GHz
\citep{hsf+11}.  There is no coincident radio source in the catalog of
M51 compact radio sources \citep{mck+07}.  A comparison with the
CARMA and SMA flux densities obtained contemporaneously indicates that
the spectral peak lies between the EVLA and CARMA bands
(Figure~\ref{fig:spectrum2}).

We began monitoring SN\,2011dh with the EVLA on June 17 UT ($\Delta
t\approx 17$ days after explosion) as part of a Rapid Response
observing program for long-term monitoring of the supernova (PI
Soderberg).  Data were collected within the wide C, X, Ku, K, and Ka
bands. Within each of these bands (except X) we selected two central
frequencies enabling spectral coverage spanning $\nu=5.0-36.0$ GHz.
Each central frequency has associated bandwidth of 0.8 to 1.0 GHz.
All observations were obtained in the (most extended) A-array.  We used
J1327+4326 to monitor the phase while the absolute flux calibration
was carried out using 3C286. Data were reduced using AIPS and the
Common Astronomy Software Applications (CASA).  We fit a Gaussian
model to the radio SN emission in each observation to derive the
integrated flux density (Table~2).  The reported flux density errors
include errors from Gaussian fitting, the map rms noise, and
systematic errors of 1\% at low frequencies (5--16 GHz) and 3\% at
high frequencies (20.5--36.0 GHz).  At this epoch, the peak of the
radio spectrum has clearly shifted to the cm-band.

Additional observations with the EVLA are on-going and are the focus
of a follow-up paper \citep{krauss11}.  In Figure~\ref{fig:spectrum2} we compare the
radio spectrum of SN\,2011dh with the newly available and unprecedented
wide bands of EVLA which enable continuous spectral coverage
from $\sim 1-40$ GHz.  SN\,2011dh represents the first SN for which
such detailed mapping of the spectrum has been possible in the EVLA
era.

\section{A Model for the Radio Emission}
\label{sec:model}

Early radio observations of SNe uniquely trace the shockwave as it races
ahead of the bulk ejecta and shock-accelerates particles in the local
circumstellar medium (CSM; \citealt{c82}).  This environment was
enriched by the progenitor star wind during the centuries leading up
to the explosion.  Through this dynamical interaction, the shockwave
accelerates CSM electrons into a power-law distribution,
$N(\gamma)\propto \gamma^{-p}$, above a minimum Lorentz factor,
$\gamma_m$.  The accelerated electrons gyrate in amplified magnetic
fields giving rise to non-thermal synchrotron emission.  In the case
of Type Ibc and cIIb supernovae, the radio emission is quenched at low
frequencies primarily due to synchrotron self-absorption (SSA), producing
a spectral turnover that defines the peak of the radio spectrum,
$\nu_p$.  The self-absorbed radio spectrum is described by
$F_{\nu}\propto \nu^{5/2}$ below $\nu_p$ and $F_{\nu}\propto
\nu^{-(p-1)/2}$ above $\nu_p$.  As shown in Figure~\ref{fig:spectrum2}, our
EVLA, CARMA, and SMA observations of SN\,2011dh on two separate epochs
are well-described by a synchrotron self-absorbed spectrum with
$p\approx 3$.  We note that the modest disagreement between the SSA
model and the measurements near the spectral peak may indicate
asphericity of the emitting region (see \citealt{krauss11} for a
detailed discussion). 

\citet{c98} showed that for radio SNe with minimal free-free absorption, 
the radius of the shockwave, $R$, and its time-averaged velocity,
$\overline{v}$, can be robustly estimated from the observed values of
$\nu_p$ and the associated peak spectral luminosity, $L_{\nu,p}$.  
As the shockwave decelerates, the optical depth to absorption
processes declines and $\nu_p$ cascades to lower frequencies; for 
the case of free expansion, we expect $\nu_p\propto t^{-1}$ and
a nearly constant peak spectral luminosity \citep{c98}.
For $p\approx 3$, the shockwave radius is given by $R\approx 3.3\times
10^{15} (\epsilon_e/\epsilon_B)^{-1/19}(L_{\nu_p,26})^{9/19}
\nu_{p,5}^{-1}~\rm cm$ where $L_{\nu_p,26}$ is normalized to
$10^{26}~\rm erg~s^{-1}~Hz^{-1}$ and $\nu_{p,5}$ is normalized to 5
GHz \citep{cf06}.  The fractions of post-shock energy density shared
by accelerated electrons and amplified magnetic fields are denoted by
$\epsilon_e$ and $\epsilon_B$, respectively, and we assume that the
radio emitting region is half of the total volume enclosed by a
spherical shockwave.

\begin{figure}
\epsscale{0.9}
\plotone{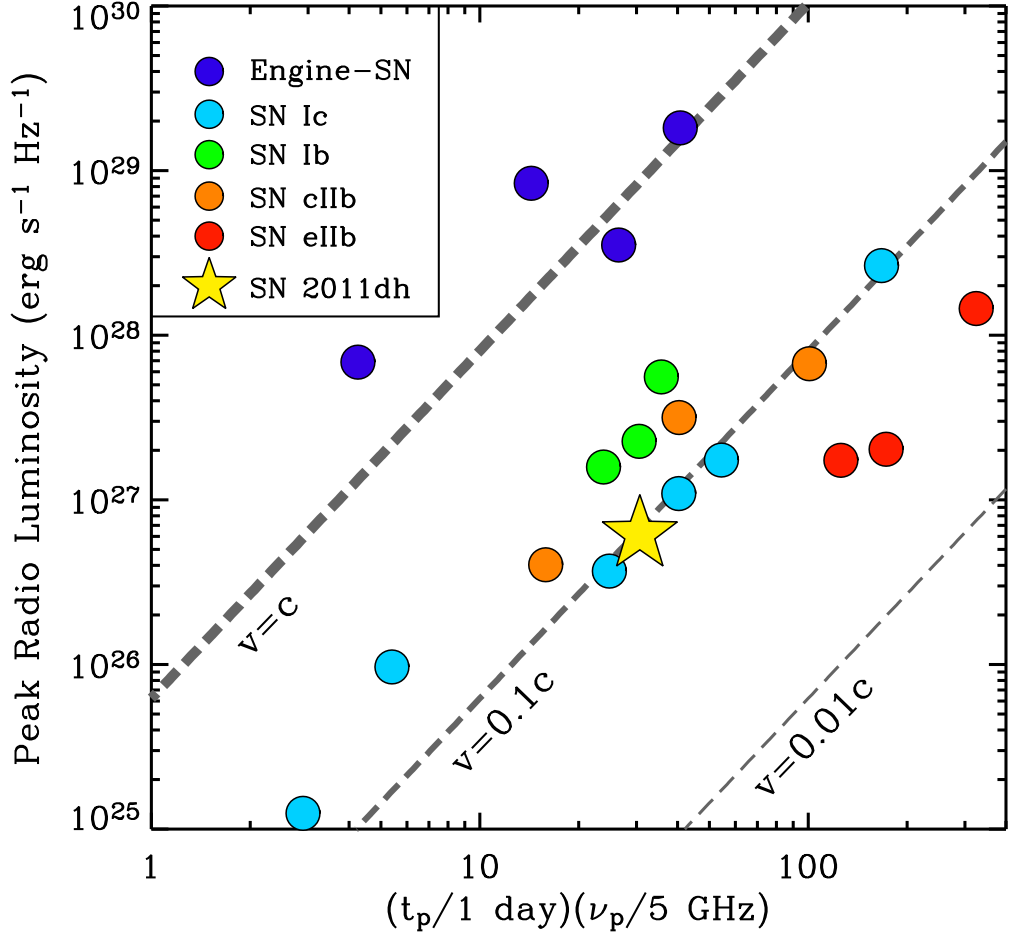}
\caption{The radio properties of various SNe are compared including SNe Ib (green) Ic (cyan), cIIb (orange), eIIb (red), and engine-driven SNe associated with nearby GRBs (blue).  The properties of the spectral 
peak can be used to derive the shockwave velocity (dashed lines).
With a blastwave velocity of $\overline{v}\approx 0.1c$, SN\,2011dh
(yellow star) is more similar to Types Ibc and cIIb than Type
eIIb. Adapted from \citet{cs10}.}
\label{fig:L_t_11dh}
\end{figure}

As shown in Figure~\ref{fig:spectrum2}, for SN\,2011dh we find
$\nu_p\approx 40$ and $\approx 11$ GHz on 2011 June 4 and 17 ($\Delta
t\approx 4$ and 17 days), respectively, with an associated peak
luminosity of $L_{\nu_p,26}\approx 6.7$ on both epochs. These
observables correspond to shockwave radii of $R\approx
(1.0,~3.7)\times 10^{15}$ cm assuming typical partition values of
$\epsilon_e=\epsilon_B=0.1$. The time averaged shockwave velocity is
thus $R/\Delta t\approx \overline{v}\approx 0.1c$ and thus a factor of $\sim 2$ faster
than the material at the optical photosphere at $\Delta t\approx 3$ days
\citep{sfc11}.  The total internal energy required to power the
observed radio signal can be estimated from the post-shock magnetic
energy density, $E=B^2 R^3/12\epsilon_B$.  As shown by \citet{cf06},
the amplified magnetic field is directly determined from the spectral
properties, $B\approx 0.70~(\epsilon_e/\epsilon_B)^{-4/19}
L_{\nu_p,26}^{-2/19}
\nu_{p,5}~\rm G$.  At $\Delta t\approx 4$ and 17 days, we find
$B\approx 4.5$ and $1.2$ G. The total internal energy is thus,
$E\approx (1.7,~6.3)\times 10^{46}~\rm erg$ for the two epochs,
respectively, by maintaining the assumption that $\epsilon_B=0.1$,
i.e. $\epsilon_{B,-1}$. The roughly linear temporal increase of
internal energy is consistent with a slightly decelerated shockwave.
As shown in Figure~\ref{fig:L_t_11dh}, the shockwave properties of
SN\,2011dh are similar to those of Type cIIb, which tend to also be
characterized by shockwave velocities of $\overline{v}\gtrsim 0.1c$
while slower shockwaves are inferred for SNe eIIb.

The progenitor mass loss rate is $\dot{M}\approx 0.39\times
10^{-5}~\epsilon_{B,-1}^{-1} (\epsilon_e/\epsilon_B)^{-8/19}
L_{\nu_p,26}^{-4/19} \nu_{p,5}^{2} t_{p,10}^2~\rm M_{\odot}~yr^{-1}$
where $t_{p,10}$ is the observed time of the spectral peak normalized
to 10 days and we have assumed a wind velocity of $v_w=1000~\rm
km~s^{-1}$.  Thus, we estimate a progenitor mass loss rate for
SN\,2011dh of $\dot{M}\approx 3\times 10^{-5}~\rm M_{\odot}~yr^{-1}$,
similar to the mass loss rates derived for SNe Ibc and cIIb and also
similar to the values observed for Galactic Wolf-Rayet stars
\citep{cgv04,cro07}, and a factor of 100 lower than the mass loss
rate inferred for SN\,1993J \citep{flc96}.  We note that in this framework, the radio data
constrain the ratio, $(\dot{M}/v_w)$, such that a variation in the
assumed value of $v_w$ shifts the mass loss rate estimate by the same
factor.  

\section{A Model for the X-ray Emission}
\label{sec:ic}

On timescales of days after explosion, the X-ray emission observed
from SNe Ibc and IIb may be dominated by a number of different
emission processes including synchrotron, thermal, and inverse
Compton (IC) scattering.  Continued energy input from a compact remnant (black hole,
magnetar) has also been invoked to explain the X-ray emission from
some events (e.g., SN\,2006aj, \citealt{paolo06,skn+06}; SN\,1979C,
\citealt{plj11}). The temporal and spectral evolution of the radio and
X-ray emission, together with their observed luminosity values,
enables these different processes to be distinguished.  Here we
consider the nature of the observed X-ray light-curve for SN\,2011dh.

The radio-to-X-ray spectral index is observed to be $\beta_{RX}\approx
-0.7$ on both June 4 and 17 UT. Thus, an extrapolation of the
optically-thin radio synchrotron spectrum with $p\approx 3$
under-estimates the X-ray flux at both epochs by a factor of $\sim
140$.  Similarly high radio-to-X-ray spectral indices are not atypical
for SNe Ibc and IIb (see \citealt{cf06} and references within).
Attributing the X-rays to synchrotron emission would require efficient
particle acceleration and a flattening of the electron distribution at
high energies.

\begin{figure}
\epsscale{1.1}
\plotone{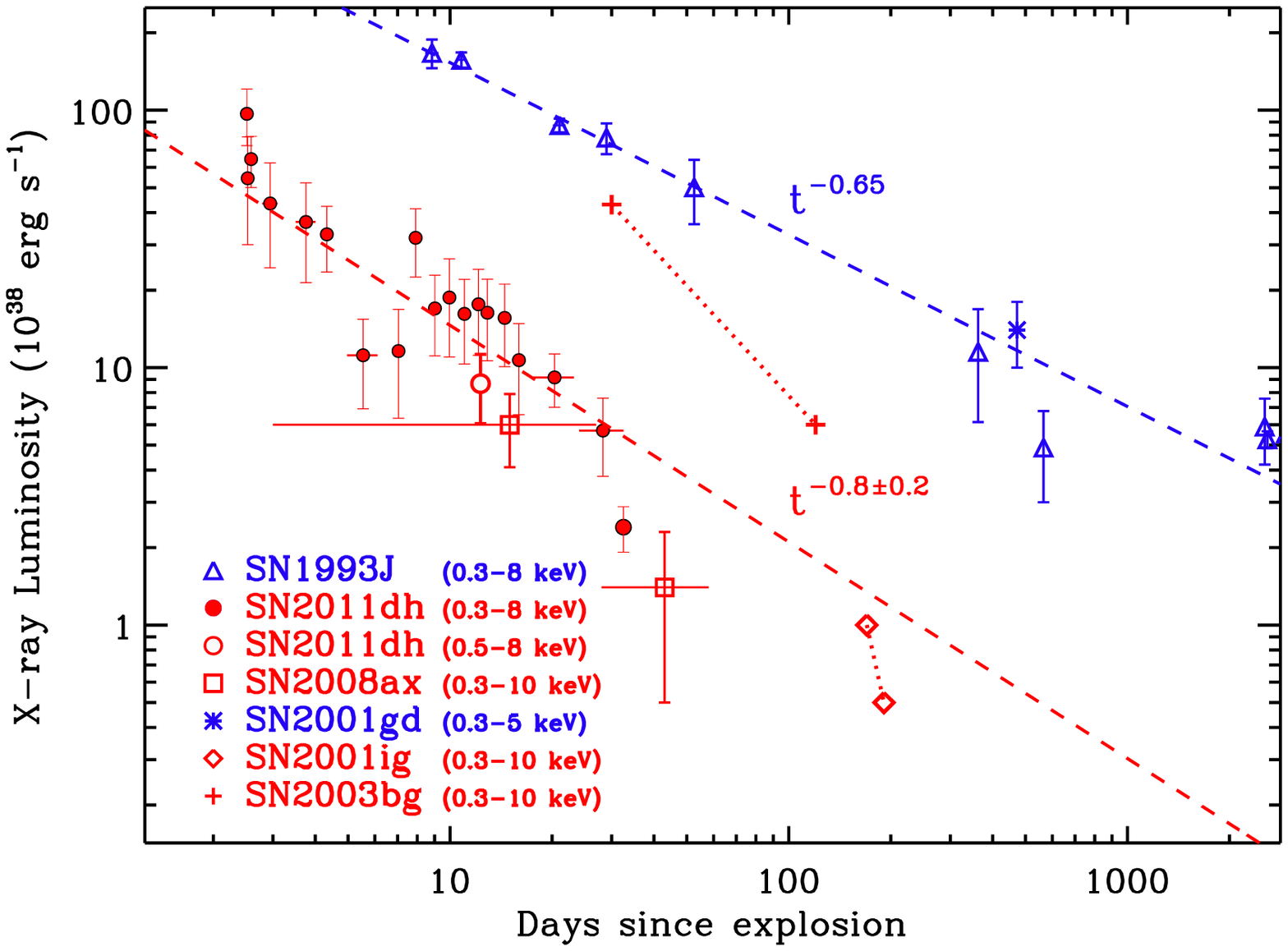}
\caption{X-ray emission from SNe IIb: SN\,1993J \citep{cdr+09}; SN\,2001gd \citep{pam+05}; SN\,2008ax \citep{rpb+09}; SN\,2001ig \citep{sr02}, and
SN\,2003bg \citep{sck+06} and SN\,2011dh (this work). Error
bars are 1$\sigma$.
For SN\,2011dh the
energy band (0.3-8) keV is used to allow a direct comparison to
SN\,1993J.  SNe of Type cIIb are shown in red while Type eIIb are shown in 
blue.  The X-ray luminosities of Type cIIb appear lower than those of Type eIIb.}
\label{Fig:2011dhvs1993J}
\end{figure}

As shown in \citet{cf03}, the thermal emission from shock heated
ejecta in the reverse shock region may give rise to strong free-free
X-ray emission.  If the material is fully ionized and in temperature
equilibrium, the expected luminosity is broadly determined by the
electron temperature and the mass loss rate of the progenitor star
with an expected linear decay in time.  We use equations (25-30) of
\citet{cf06} with $\overline{v}\approx 0.1c$ and
$\dot{M}\approx 3\times 10^{-5}~\rm M_{\odot}~yr^{-1}$ to
estimate the electron temperature\footnote{We note that \citet{cf06} adopt the notation 
$A_*\equiv  A/(5\times 10^{11} \rm g cm^{-1})$ where $A=\dot{M}/4\pi v_w$.}.  The expected free-free X-ray
luminosity for SN\,2011dh 
is then $L_{ff}\approx 2.5\times 10^{37}~\rm erg~s^{-1}$ on June 4
UT.  This is
a factor of $\sim 200$ lower than the {\it Swift}/XRT measurement at
this epoch (Figure~\ref{Fig:2011dhvs1993J}).

We next consider an inverse Compton scattering model in which the
optical photons associated with both envelope cooling emission and the
radioactive decay of $^{56}$Ni are upscattered to the X-ray band by
radio emitting electrons \citep{cfn06}. In this scenario, the X-ray
decay should track the optical evolution through the cooling envelope
decay to the re-brightening due to $^{56}$Ni decay.  The early optical
emission indicates an average luminosity of $L_{\rm bol}\sim {\rm
few}\times 10^{42}~\rm erg~s^{-1}$ and a minimum near $\Delta t\approx
5$ days.  The X-ray light-curve suggests a similar minimum near this
epoch and an overall decay of roughly $L_X\propto t^{-1}$
(Figure~\ref{Fig:2011dhvs1993J}). Adopting the formalism of
\citet{cf06} and assuming the relativistic electron population extends
down to ($\gamma_m=1$), the predicted IC emission is $L_{IC}\approx
2.9\times 10^{36}~\epsilon_{B,-1} \dot{M}_{-5}
(\epsilon_e/\epsilon_B)^{11/19} (v/0.1c)^{-1} L_{\rm bol,42} \Delta
t_{d,10}^{-1}~\rm erg~s^{-1}$ where $\dot{M}_{-5}$ is the mass loss
rate normalized to $10^{-5}~\rm M_{\odot}~yr^{-1}$ and we maintain the
assumption of $v_w=10^3~\rm km~s^{-1}$.  Here, $\Delta t_{d,10}$ is
the time since explosion normalized to 10 days.

We find that the X-ray emission may be attributed to inverse Compton
emission if the assumption of equipartition is relaxed and
$(\epsilon_e/\epsilon_B)\approx 30$ with $\epsilon_B\approx 0.01$.
This deviation from equipartition implies modest adjustments to our
physical parameters estimates, including a factor of $\sim 2$ increase
in the mass loss rate and a minimal increase in the inferred total 
internal energy of the radio emitting material.  The modified values
for the mass loss rate and magnetic field are $\dot{M}_{-5}\approx 6$
and $B\approx 2.2$ G, respectively.  A prediction of this model is
that the X-ray light-curve will decay more steeply following the
optical SN peak \citep{cf06}, which may be suggested by our {\it
Chandra} measurement on July 3 UT.

\section{Constraints on the Progenitor Size}
\label{sec:prog}

The early optical emission from SNe is dominated by the adiabatic
cooling of envelope material following the breakout of the shockwave
through the stellar surface \citep{eb92}.  This component cascades
through the optical band in the first few days following explosion.
The radius and temperature associated with this component may be
roughly approximated as thermal (although see \citealt{ns10} for a
more comprehensive discussion of the spectral evolution) and used
together with estimates for the bulk SN parameters (including the
ejecta kinetic energy, $E_K$, and mass, $M_{\rm ej}$) leads to a
determination of the progenitor radius, $R_*$.  Such techniques have
been used to derive the progenitor radius of SNe 1987A \citep{eb92},
1999ex and 2008ax \citep{cs10}, 2008D \citep{cf08,sbp+08}, in addition
to SN\,2006aj associated with XRF\,060218 \citep{campana06,wmc07}. Here
we adopt the formalism of \citet{cf08} in which the temperature of the
photosphere is given by $T_{\rm ph}\approx 7800~E_{K,51}^{0.03}~M_{\rm
  ej,\odot}^{-0.04} R_{*,11}^{0.25} \Delta t_{d}^{-0.48}~\rm K$ and
the photospheric radius is $R_{\rm ph}\approx 3\times
10^{14}~E_{K,51}^{0.39}~M_{\rm ej,\odot}^{-0.28} \Delta
t_{d}^{0.78}~\rm cm$.  Here we have normalized $E_{K,51}$ to $10^{51}$
erg, $M_{\rm ej,\odot}$ to $M_{\odot}$ and $R_{*,11}$ to $10^{11}$
cm. Within this framework, it is assumed that the density and pressure
profiles of the envelope are consistent with those of \citet{mm99},
and that the bolometric luminosity decays as $L_{\rm ph}\propto \Delta
t_{d}^{-0.34}$.

We model the first few optical observations ($\Delta t\lesssim 5$
days) including the initial detection of $L_{\rm ph}\approx 10^{42}~\rm
erg~s^{-1}$ on June 1.191 UT \citep{agy+11}.  \citet{sfc11} report a
photospheric velocity of $v_{ph}\approx 17,600~\rm km~s^{-1}$ at
$\Delta t\approx 3$ days after explosion implying a ratio of
$(E_{K,51}/M_{\rm ej,\odot})\sim$ few.  We estimate $R_{\rm ph}\approx
5\times 10^{14}~\rm cm$ and $T_{\rm ph}\approx 8000$ K at $\Delta
t\approx 1$ day.  This is roughly consistent with the photospheric
temperature derived from optical spectroscopy
\citep{agy+11}, and implies a compact progenitor since $T_{\rm
ph}\propto R_{*,11}^{0.25}$.  The high luminosity of the initial
detection requires a ratio, $E_K/M_{\rm ej}\sim $few, in line with the
ratio implied by the high photospheric velocity.  Thus, the early
optical emission points to a compact progenitor, similar to those of
SNe Ibc and cIIb, and consistent with the earlier report by
\citet{agy+11} and the conclusions of \citet{murphy11} based
on a study of the SN\,2011dh environment.

We note, however, that the observed optical decay, $L\propto \Delta
t_d^{-1}$, is significantly steeper than the model prediction.  This
may point to an irregular density profile near the stellar surface,
perhaps associated with an unusual mass loss ejection in the final
stage of the progenitor's evolution.  Such an irregular density profile 
may affect some of the conclusions that we derived above based on the
canonical model and this will be the focus of a
future publication. 

\section{Shock Breakout X-ray Emission}
\label{sec:hurley}

For compact progenitors such as SN\,2011dh, the breakout pulse may be
detectable at X-ray and gamma-ray energies (e.g., SN\,2008D;
\citealt{sbp+08}).  To this end, we searched for evidence of a high
energy pulse associated with the shock breakout of SN\,2011dh using
data collected by the {\it Swift} Burst Alert Telescope (BAT), the
Monitor of All-sky X-ray Image (MAXI) camera attached to the Japanese
Experiment Module, and the Interplanetary Network (IPN: Mars Odyssey,
Konus-Wind, RHESSI, INTEGRAL (SPI-ACS), {\it Swift}/BAT, Suzaku, AGILE, and
Fermi/GBM; we note that MESSENGER was in superior conjunction and
off during this period).

\begin{figure}
\plotone{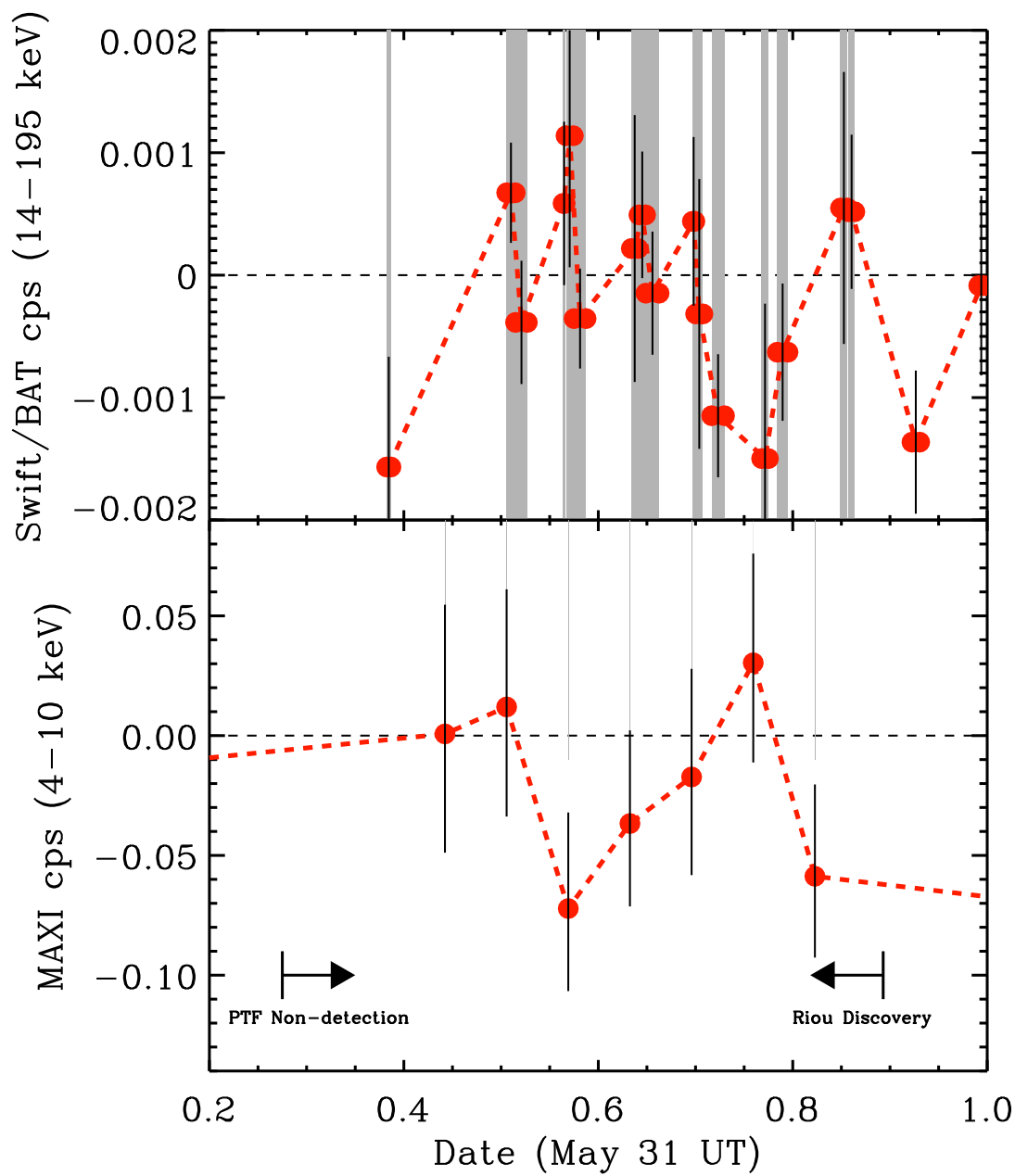}
\caption{Swift/BAT (15-195 keV) and MAXI (4-10 keV) observations of
  the SN\,2011dh field during the
  estimated explosion date range bounded by the discovery by Riou and
  pre-discovery limit (PTF) marked with arrows, May 31.275-31.893 UT
  \citep{agy+11}. No source is detected at the position of the SN during the individual ($\sim 1000$ sec) BAT pointings or ($\sim 80$ sec) MAXI scans.  The time ranges when the SN was in the FOV of the instruments are shown in 
grey.}
\label{fig:gray}
\end{figure}

The position of SN\,2011dh was observed by {\it Swift}/BAT over many
pointings during the estimated explosion date range, May 31.275-31.893
UT.  No gamma-ray emission was detected from the SN.  We compile the
{\it Swift}/BAT observations in Figure~\ref{fig:gray}.  Each pointing
had a duration of $\sim 500-1000$ sec, and the total time on source
was 11148 sec or 20\% of the full explosion date range.  We estimate
the count rate at the SN position and also in the background in each
pointing and note that the sensitivity varies as a function of the
off-axis angle, ranging from 10 to 57 degrees.  Assuming a power law
spectrum for the breakout flux with a photon index of $\Gamma=2$, we
infer $5\sigma$ upper limits on the gamma-ray emission from the SN of
$F_{\gamma}\lesssim (1.1-3.9)\times 10^{-9}~\rm erg~cm^{-2}~s^{-1}$
(15-195 keV) for each individual pointing.  The combined upper limit
is $F_{\gamma}\approx 2.8\times 10^{-10}~\rm erg~cm^{-2}~s^{-1}$ for
the same energy range.  At the distance of M51, this limit corresponds
to a luminosity of $L_{\gamma}\lesssim 2.4\times 10^{42}~\rm
erg~s^{-1}$.  We further note that no emission was detected by {\it
  Swift}/BAT in the direction of the SN during spacecraft slews
throughout this time period; the BAT Slew Survey \citep{grindlay07} places a
$\sim 4\sigma$ upper limit within energy
range, (15-50) keV. 

MAXI \citep{mku+09} is an X-ray all sky monitor that scans most of the
sky every 92 minutes with its Gas Slit Cameras (GSC;
\citealt{mns+11}).  Unfortunately during the estimated period of the
SN breakout, the direction of M51 was covered only by a camera with
high background, resulting in less stringent limits than usual.  The
X-ray image taken at every scan transit ($\sim 80$ sec), encoded in
one dimension by the detector position and the other by transit
timing, was fitted to the PSF with the fixed source position to
evaluate the source count rate.  The energy range of (4-10) keV was
chosen to maximize the signal-to-noise ratio.  Then the count rate was
converted to energy flux assuming a power-law model with a photon
index of $\Gamma=2$ resulting in a conversion factor of $F_X\approx
1.1\times 10^{-8}~\rm erg~cm^{-2}~s^{-1}~cps^{-1}$. We derive an average
$5\sigma$ upper limit of $L_X\lesssim 3.9\times 10^{42}~\rm
erg~s^{-1}$.  In comparison, the observed breakout pulse from
SN\,2008D ($d\approx 30$ Mpc) had a luminosity $\sim 25$ times
higher ($L_X\approx 10^{44}~\rm erg~s^{-1}$; 0.3-10 keV) with a
duration of $\sim 5$ min.

While {\it Swift}/BAT and MAXI both have a limited temporal coverage, 
the IPN is full sky with temporal duty cycle of nearby 100\%; it is
sensitive to hard X-ray emission in the energy range, (25-150) keV
\citep{hga+10}. Within a two-day window centered on the 
estimated explosion date, a total of four triggered bursts were
detected by the IPN.  All four were confirmed through observations by
multiple instruments or spacecrafts and thus could be localized.
These detections include Fermi/GBM (3 bursts), Konus (1), INTEGRAL (1) and
{\it Swift}/BAT (1). In all cases the localizations were statistically
{\it inconsistent} at a $\gtrsim 3\sigma$ level with the position of
SN\,2011dh.  Thus we find no evidence for a detected gamma-ray
transient in coincidence with SN\,2011dh and adopt an upper limit
of $F_{\gamma}\lesssim 6\times 10^{-7}~\rm erg~cm^{-2}~s^{-1}$ corresponding to
an energy of $E_{\gamma}\lesssim 5\times 10^{45}$ erg for SN\,2011dh.  

\subsection{A Comparison to Breakout Predictions}
\label{sec:boaz}

Here we compare theoretical predictions for a prompt breakout pulse
with the available X-ray and $\gamma-$ray limits.  We consider a model
in which the progenitor is compact, $R_{*,11}\approx 1$, and embedded
in a stellar wind with $\dot{M}_{-5}\approx 6$.  In the standard SN
Ibc model of \citet{cf06}, also applicable to SNe cIIb, the shockwave
radius evolves as $R\propto t^m$ with $m\approx 0.9$.  Extrapolating
back to the breakout radius, the shockwave speed at the stellar
surface is $v_0\approx 0.3c$ and we define $\beta_0\approx
0.3\beta_{-0.5}$ as the the shock velocity (divided by $c$) at this time.  
In the case of such fast shock velocities, there are deviations from
thermal equilibrium resulting in high energy emission \citep{kbw10,ns10,skw11}.

The optical depth at the stellar surface is $\tau=\kappa \dot{M}/4\pi
v_w r \approx 2\dot{M}_{-5} R_{11}^{-1}$ where we adopt
$\kappa=0.4~\rm cm^2~g^{-1}$ and $R_{11}$ is the shock radius
normalized to $10^{11}$ cm.  The shock breakout will occur outside of
the star, within the stellar wind at a radius, $R_{\rm
br}=\beta_0\kappa \dot{M}/4\pi v_w\approx 6\times
10^{10}~\dot{M}_{-5}\beta_{-0.5}\rm~cm$ as long as $R_*<R_{br}$.  In
the case of SN\,2011dh, we therefore predict a wind breakout at
radius, $R_{br}\approx 4\times 10^{11}$ cm.  The energy associated
with breakout is $E_{br}\approx 3\times
10^{43}~\dot{M}_{-5}^2\beta_{-0.5}^3$ \citep{ksw11}.  On the breakout
timescale, a radiative collisionless shock forms and an additional
comparable or larger amount of energy is emitted \citep{ksw11}.
Adopting our parameters for SN\,2011dh, we predict $E_{br}\gtrsim
10^{45}$ erg. During the transition from a radiation mediated shock to
collisionless, the temperature rises steadily from $\gtrsim$ keV to
$\gtrsim 100$ keV and we roughly estimate the spectrum to be $\nu
F_{\nu}\sim$ constant. 

The predicted breakout energy is within the detectability range of the
{\it Swift}/BAT and MAXI upper limits spanning (4-195) keV, however,
both offer only limited temporal coverage.  We estimate the rise time
of the breakout pulse to be $t_{br}\approx R_{br}/v_0\approx 1$ min.
Efficient emission from the collisionless shock continues beyond this
time, and the full duration of the pulse is necessarily longer.  Given
the large gaps in the {\it Swift}/BAT and MAXI coverage (see
Figure~\ref{fig:gray}), especially toward the beginning of the
explosion date range, we speculate that {\it Swift} and MAXI missed
the breakout pulse.  The IPN upper limit is a factor of a few above
the predicted energy and thus consistent with this breakout model.
Finally we note that if the progenitor was more extended with a
radius, $R_{*}\approx 10^{13}$ cm, a stronger pulse of
$E_{br}\sim 3\times 10^{47}~R_{*,13}^2~$erg is expected at lower
frequencies, $h\nu\lesssim 300\rm eV$. Such a pulse is not constrained
by the hard X-ray/gamma-ray observations presented here.

\section{Conclusions}
\label{sec:conc}

We present multi-wavelength follow-up observations of SN\,2011dh
spanning the radio, millimeter, X-ray, and gamma-ray bands and
obtained within the first few seconds to weeks following the
explosion.  The X-ray light-curve for SN\,2011dh is perhaps the
best-sampled to date for a SN IIb and suggests that X-ray properties
(luminosity, spectral evolution) may further distinguish compact and
extended progenitors.  Using the newly available wide bands of the
EVLA, we show that the radio and millimeter data are consistent with a
synchrotron self-absorbed spectrum, while the X-ray emission may be
understood as inverse Compton upscattering of optical SN photons
assuming a non-negligible deviation from equipartition of the shock partition fractions.
These data offer unique diagnostics (shockwave velocity,
$\overline{v}\approx 0.1c$) and point to a compact progenitor
star. Through modeling of the early optical emission we find that the
photosphere is characterized by a low temperature, $T_{ph}\approx
8000$ K, and the physical parameters of the ejecta are constrained to
be $E_{K,51}/M_{\rm ej,\odot}\sim $ few.  These properties point to a
compact progenitor, $R_*\approx 10^{11}$ cm, however, we find that the
fast decay of the early optical emission at $\Delta t\lesssim 5$ day
is incompatible with canonical cooling envelope models and may suggest
an irregular ejecta profile.  A compact progenitor size appears
inconsistent with the extended radius ($R_*\sim 10^{13}$ cm) of the
coincident yellow supergiant identified in pre-explosion HST imaging,
suggesting an unusual density profile for the outer layers of the progenitor star.
It may be possible to explain the coincident object as a binary companion,
however, it is estimated that the yellow supergiant phase lasts just $\sim 3000$ years
\citep{dmm+09} making this scenario improbable.  At the same time, the rarity of 
yellow supergiants makes a positional coincidence unlikely, suggesting
the supergiant is indeed related to the progenitor system.  

If the supergiant was characterized by a low mass envelope and a
low density stellar wind it may be possible to reconcile the radio
and X-ray observations with the putative progenitor.  \citet{georgy}
recently suggested that the mass loss rates for yellow supergiants 
have been theoretically over-estimated.  However, the early cooling
envelope emission require a substantially lower density envelope
then those associated with supergiants; in this scenario, ejecta asymmetries
may be the only way to reconcile the early optical data with the
coincident supergiant as the progenitor.

We conclude that SN\,2011dh is likely a member of the Type cIIb
class of core-collapse explosions.  Our long-term EVLA monitoring
observations will reveal if the radio light-curves are modulated   (see \citealt{krauss11} for details),
indicative of a variable and/or episodic mass loss history in the
decades leading up to explosion -- an observational characteristic
shared by other SNe cIIb.   In addition, our long-term monitoring with 
the Very Long Baseline Interferometer (VLBI) will provide an
direct measurement of the shock radius and, in turn, an independent constraint
on the shock partition fractions \citep{marti11,biet12}.  Future observations of the explosion site with {\it   HST} imaging will
reveal whether the supergiant has disappeared, thereby directly linking it to
the explosion.

Finally, we used the parameters derived above for the shockwave and
progenitor to estimate that the shock breakout pulse from SN\,2011dh
was detectable in the X-ray and gamma-ray bands with current
satellites. We attribute the lack of a detection as likely due to limited
temporal/spatial coverage during the estimated explosion date.
Looking forward, sensitive and wide-field X-ray experiments such as
LOBSTER and JANUS will regularly discover shock breakout emission from
similarly nearby and compact SN progenitors (see
e.g.,~\citealt{sgb+10} for a discussion). Such detections will not
only provide information on the progenitor size since the
explosion date estimates will enable statistically significant
searches for gravitational wave and neutrino counterparts
\citep{ott09,ksw11}.  Furthermore, these prompt discoveries will
enable rapid low-frequency follow-up with sensitive radio and
millimeter facilities thanks to the advent of EVLA, the Atacama
Large Millimeter Array (ALMA), ASKAP, and MeerKAT.

We thank Philip Massey, Edo Berger, Ryan Foley, Maria Drout and
Robert Kirshner for useful conversations.  KH is grateful for IPN
support from NASA grants NNX10AI23G, NNX09AU03G, NNX07AR71G, and
NNX10AR12G.  BK and EML are supported by NASA through Einstein Postdoctoral
Fellowship awarded by the Chandra X-ray Center, which is operated by
the Smithsonian Astrophysical Observatory for NASA under contract
NAS8- 03060.  Support for CARMA construction was derived from the
Gordon and Betty Moore Foundation, the Kenneth T. and Eileen L. Norris
Foundation, the James S. McDonnell Foundation, the Associates of the
California Institute of Technology, the University of Chicago, the
states of California, Illinois, and Maryland, and the National Science
Foundation. Ongoing CARMA development and operations are supported by
the National Science Foundation under a cooperative agreement, and by
the CARMA partner universities.  The SMA is a joint project between
the Smithsonian Astrophysical Observatory and the Academia Sinica
Institute of Astronomy and Astrophysics, and is funded by the
Smithsonian Institution and the Academia Sinica.  The National Radio 
Astronomy Observatory is a facility of the National Science Foundation
operated under cooperative agreement by
Associated Universities, Inc.

\bibliographystyle{apj1b}

%\bibliography{journals_apj,refs}

\clearpage

\clearpage

\clearpage

\clearpage

\clearpage

\begin{deluxetable}{lcccl}
\tablecaption{X-ray Observations of SN\,2011dh}
\tablewidth{0pt}
\tablehead{
\colhead{Date} & 
\colhead{Time Range} &
\colhead{Unabsorbed Flux\tablenotemark{1}} &
\colhead{Error} & 
\colhead{Satellite} \\
\colhead{(UT)}  &  
\colhead{(days)} & 
\colhead{($\rm erg~cm^{-2}~s^{-1}$)} &
\colhead{($\rm erg~cm^{-2}~s^{-1}$)} &  
\colhead{} \\
}
\startdata
       June 3.512   &   0.004 &  $1.12\times 10^{-12}$ &  $2.8\times 10^{-13}$   &     {\it Swift}/XRT \\
       June 3.526   &   0.003 &  $6.28\times 10^{-13}$ &  $2.8\times 10^{-13}$   &     {\it Swift}/XRT \\
       June 3.585   &   0.007 &  $7.45\times 10^{-13}$ &  $1.7\times 10^{-13}$   &     {\it Swift}/XRT \\
       June 3.943   &   0.152 &  $5.01\times 10^{-13}$ &  $2.2\times 10^{-13}$   &     {\it Swift}/XRT \\
       June 4.753   &   0.252 &  $4.25\times 10^{-13}$ &  $1.8\times 10^{-13}$   &     {\it Swift}/XRT \\
       June 5.326   &   0.033 &  $3.80\times 10^{-13}$ &  $1.1\times 10^{-13}$   &     {\it Swift}/XRT \\
       June 6.536   &   0.571 &  $1.29\times 10^{-13}$ &  $4.9\times 10^{-14}$   &     {\it Swift}/XRT \\
       June 8.042   &   0.181 &  $1.34\times 10^{-13}$ &  $6.0\times 10^{-14}$   &     {\it Swift}/XRT \\
       June 8.911   &   0.253 &  $3.69\times 10^{-13}$ &  $1.1\times 10^{-13}$   &     {\it Swift}/XRT \\
      June 10.020   &   0.301 &  $1.96\times 10^{-13}$ &  $6.8\times 10^{-14}$   &     {\it Swift}/XRT \\
      June 10.954   &   0.165 &  $2.16\times 10^{-13}$ &  $8.9\times 10^{-14}$   &     {\it Swift}/XRT \\
      June 12.027   &   0.370 &  $1.87\times 10^{-13}$ &  $6.7\times 10^{-14}$   &     {\it Swift}/XRT \\
      June 13.127   &   0.179 &  $2.03\times 10^{-13}$ &  $7.4\times 10^{-14}$   &     {\it Swift}/XRT \\
      June 13.887   &   0.200 &  $1.89\times 10^{-13}$ &  $6.6\times 10^{-14}$   &     {\it Swift}/XRT \\
      June 15.493   &   0.601 &  $1.80\times 10^{-13}$ &  $6.4\times 10^{-14}$   &     {\it Swift}/XRT \\
      June 16.969   &   0.136 &  $1.23\times 10^{-13}$ &  $4.8\times 10^{-14}$   &     {\it Swift}/XRT \\
      June 21.345   &   2.848 &  $1.06\times 10^{-13}$ &  $2.5\times 10^{-14}$   &     {\it Swift}/XRT \\
      June 29.278   &   4.227 &  $6.57\times 10^{-14}$ &  $2.2\times 10^{-14}$   &     {\it Swift}/XRT \\ 
\hline 
      June 12.300   &   0.115 &  $1.00\times 10^{-13}$ &   $3.0\times 10^{-14}$  &      Chandra/ACIS\tablenotemark{\dagger} \\
      July 3.400    &   0.115 &  $2.75\times 10^{-14}$ &  $5.5\times 10^{-15}$   &      Chandra/ACIS \\

\enddata
\tablenotetext{1}{Energy range, (0.3-8) keV.}
\tablenotetext{\dagger}{Observation reported by \citet{poo+11} within energy range (0.5-8) keV.}
\label{tab:xray}
\end{deluxetable}

\clearpage

\begin{deluxetable}{lcccl}
\tablecaption{Radio and Millimeter Observations of SN\,2011dh}
\tablewidth{0pt}
\tablehead{
\colhead{Date} & 
\colhead{Central Frequency} &
\colhead{Flux Density} &
\colhead{Error} & 
\colhead{Telescope} \\
\colhead{(UT)}  &  
\colhead{(GHz)} & 
\colhead{(mJy)} &
\colhead{(mJy)} &  
\colhead{} \\
}
\startdata
June 4    & 22.5 & 2.68   & 0.10 & EVLA\tablenotemark{\dagger} \\
\nodata   & 107  & 4.5    & 0.3 & CARMA \\
\nodata   & 230  & $\lesssim 3.5$    & \nodata & CARMA  \\
\nodata   & 230  & 3.6    & 0.9 & SMA \\
\hline
June 17 & 5.0  &  2.430 & 0.037 & EVLA \\
\nodata & 6.8  &  4.090 & 0.048 & EVLA \\
\nodata & 8.4  &  5.535 & 0.057 & EVLA \\
\nodata & 13.5 &  6.805 & 0.072 & EVLA \\
\nodata & 16.0 &  6.721 & 0.070 & EVLA \\
\nodata & 20.5 &  6.472 & 0.195 & EVLA \\
\nodata & 25.0 &  5.127 & 0.155 & EVLA \\
\nodata & 29.0 &  4.603 & 0.140 & EVLA \\   
\nodata & 36.0 &  3.473 & 0.108 & EVLA \\
\enddata
\tablenotetext{\dagger}{Observation reported by \citet{hsf+11}.}
\label{tab:radio}
\end{deluxetable}

\end{document}